\documentclass[prb,twocolumn,showpacs,floatfix]{revtex4}
\usepackage{graphicx}
\usepackage{subfigure}
\usepackage{dcolumn}
\usepackage{bm}

\newcommand{\ColorOnline}{(Color online) }

\newlength{\graphiclength}

\begin{document}


\title{Strong enhancement of transport by interaction on contact links}

\author{Dan Bohr}
 \affiliation{MIC, Department of Micro- and Nanotechnology, NanoDTU,
 Technical University of Denmark, DK-2800 Kgs.~Lyngby, Denmark}%
\author{Peter Schmitteckert}%
 \affiliation{TKM, Institut f\"{u}r Theorie der Kondensierten Materie,
 Universit\"{a}t Karlsruhe, %
 D-76128 Karlsruhe, Germany.}%

\date{\today}

\begin{abstract}
Strong repulsive interactions within a one dimensional Fermi
system in a two-probe configuration normally lead to a reduced
off-resonance conductance. We show that if the
repulsive interaction extends to the contact regions, a strong
increase of the conductance may occur, even for systems where one
would expect to find a reduced conductance. An essential
ingredient in our calculations is a momentum-space representation
of the leads, which allows a high energy resolution.  Further, we
demonstrate that these results are independent of the high-energy
cutoff and that the relevant scale is set by the Fermi velocity.
\end{abstract}

\maketitle %

\section{Introduction}
Constructing a transport theory for  strongly correlated systems
is one of the major challenges of condensed matter physics. Even
though many interesting ideas have been proposed during recent
years, no consensus has yet emerged as to the general validity and
applicability of the various schemes. With this state of affairs,
it is of high importance to establish reliable benchmarks for
simple model systems, which then can be used to validate
new approaches.

Recently we presented a novel method for calculating linear response
conductance\cite{Bohr_Schmitteckert_Woelfle:2006} using the density
matrix renormalization group (DMRG) method\cite{White:1992}. A major
challenge in this work consisted in minimizing finite size effects,
which was achieved via modified boundary conditions. In this paper we
circumvent these technical problems by reformulating the leads in
momentum space. This approach enables us to (i) reach a much
higher energy resolution ($\sim 10^{-5}$) and (ii) allows for a
greater flexibility in the choice of discretization schemes.

In two recent papers Mehta and Andrei \cite{Mehta_Andrei:2006,Mehta_Andrei:2007}
presented nonequilibrium Bethe ansatz results for the interacting
resonant-level model (IRLM), where a single spinless level is coupled to a
left and a right lead both via a tunneling and a
density-density interaction term. However, their work currently excludes
the regime of resonant tunneling--i.e., the regime where the conductance is close to unity.\cite{Mehta_Chao_Andrei:2007}

In this work we study the linear conductance of the IRLM on a
lattice to provide a benchmark for the universal properties of the
model. In addition, we present results for an extended model, where
the central region consists of three sites, with a similar
interaction as in the IRLM model. As we will show, this model
displays the same qualitative behavior as the IRLM. It should be
noted that despite its simplicity, the IRLM captures much of
the physics of transport through
an arbitrary interacting nanostructure provided that only a single
level is close to the Fermi energy of the leads, with all other
levels  well separated and outside the voltage window within which
the transport takes place.
For perfect coupling the IRLM model corresponds to the one-dimensional
model studied by Vasseur \emph{et~al.},\cite{Vasseur_Weinmann_Jalabert:2006} 
and Molina \emph{et al.},\cite{Molina_Weinmann_Jalabert_Ingold_Pichard:2003}
obtained by restricting their nanostructure to a single site.
Using the embedding method they showed that smoothing
the ramp of interaction for perfect contacts can compensate
for the decrease of transmission
due to interaction on the nanostructure.
Here we go far beyond the energy resolution attained in previous work and
show that interaction on the contact links can lead to strong renormalization effects,
enhancing transport beyond the noninteracting system.

\section{Method and models}
We use the DMRG method to evaluate the linear response conductance of the
interacting nanostructure. In previous
work\cite{Bohr_Schmitteckert_Woelfle:2006} the leads were modeled in
\emph{real-space} by nearest-neighbor hopping chains. While simple
to implement there are several drawbacks of this method, most
prominently the need for ``damped boundary conditions'' and the
resulting problem of trapping of fermions on the Wilson chain (the
damped region).\cite{Bohr_Schmitteckert_Woelfle:2006}

\begin{figure}[tb]
\begin{center}
    \setlength{\graphiclength}{0.475\textwidth}
    \includegraphics[width=\graphiclength]{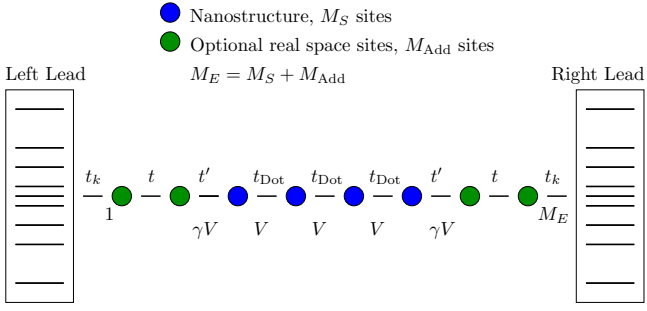}
    \caption{\ColorOnline Schematics of the nanostructure extended by real
    space sites and attached to momentum-space leads.}\label{Fig:SiteLayout}
\end{center}
\end{figure}

In the present work we introduce a setup where the leads are described
in \emph{momentum space}. Specifically, a short part of the lead
close to the nanostructure is represented in real-space, accounting
for local (i.e., high energy) physics, while further away from the
nanostructure the lead is represented in momentum space; see
Fig.~\ref{Fig:SiteLayout}. Since the low-energy modes of the
momentum leads are now directly coupled to the extended structure
(the nanostructure plus additional real-space sites), as illustrated in
Fig.~\ref{Fig:SiteLayout}, the trapping of fermions on the low
energy sites\cite{Bohr_Schmitteckert_Woelfle:2006} is avoided and no
scaling sweeps are needed. This enables much higher energy
resolution, and in the current work we resolve resonances of widths
${\mathcal O}(10^{-5})$.

By virtue of the momentum representation of the leads the
discretization scheme can be chosen arbitrarily to suit the problem
at hand. In the present work we use a logarithmic discretization to
cover a large energy range, while switching to a linear
discretization for the lowest-energy states in order to describe
Fermi-surface physics accurately. The linear discretization 
on the low-energy scale allows for a better representation of the low-energy physics
relevant for transport properties--i.e., excitations created by $\eta$.

The models considered in this work are the IRLM and the natural extension of this model to resonant linear 
chains, defined by the Hamiltonians
\begin{eqnarray}
    H_{RS} &=& \sum_{j\in S} \mu_g \hat{c}_j^\dagger  \hat{c}^{}_j
       \,-\, \sum_{j,j-1\in S_E} \big(t_j  \hat{c}_j^\dagger \hat{c}^{}_{j-1}+ \text{h.c.}\big)\nonumber\\
    && +\, \sum_{j,j-1\in S_E} V_j \Big( \hat{n}_j-\frac 12\Big) \Big( \hat{n}_{j-1} - \frac 12\Big),\\
    H_{MS} &=& \sum_{i\in L,R} \epsilon_i \hat{c}_i^\dagger  \hat{c}^{}_i,\\
    H_{T} &=& -\Big( \sum_{k\in L}t_k\hat{c}_{k}^\dagger  \hat{c}^{}_{1}
            \,+\,   \sum_{k\in R} t_k\hat{c}_{k}^\dag  \hat{c}^{}_{M_E} \Big) \,+\, \text{h.c.},
\end{eqnarray}
where $\hat{c}^\dag_\ell$ and $\hat{c}^{}_\ell$ are the (spinless)
fermionic creation and annihilation operators at site $\ell$,
$\hat{n}_\ell = \hat{c}^\dag_\ell \hat{c}^{}_\ell$. $H_{RS}$,
$H_{MS}$, and $H_T$ denote real-space, momentumspace, and tunneling
between real- and momentum-space Hamiltonians, respectively. The
symbols $S$ and $S_E$ denote the nanostructure and the extended
nanostructure (the full real-space chain), respectively.  The
indices $1$ and $M_E$ denote the first and last site in $S_E$. The
general setup and the specific values of the hopping matrix elements
$t_j$ and the interactions $V_j$ are indicated in
Fig.~\ref{Fig:SiteLayout}, and note specifically the interactions on
the contact links, $\gamma V$. The coupling $t_k$ of the extended real-space
structure to the momentum leads is chosen in such a way that in the case
of a cosine band it corresponds
to a nearest-neighbor hopping chain in real-space with a 
hopping parameter of $t$. In the following we measure all energies in units of $t=1$.

%

For a single-site nanostructure and $\gamma=1$ this model reduces
to the IRLM. The properties of the leads are defined by the
band structure $\epsilon_k$, which can take any form. In this work
we use either the cosine band, $\epsilon_k= -2 \cos(k)$, or the
linear band, $\epsilon_k = 2 k$. $D$ is a cutoff parameter
such that the Fermi velocity $v_{\text{F}}=2$ is
kept constant in all work presented here, and the band ranges between energies $-D$ and $D$.
Throughout this work we use the notion of ``contact interaction'' for interaction on
the link between the nanostructure and the leads.

\begin{figure}[tb]
\begin{center}
    \setlength{\graphiclength}{0.475\textwidth}
    \subfigure[]{\label{G_0.01}
    \includegraphics[width=\graphiclength]{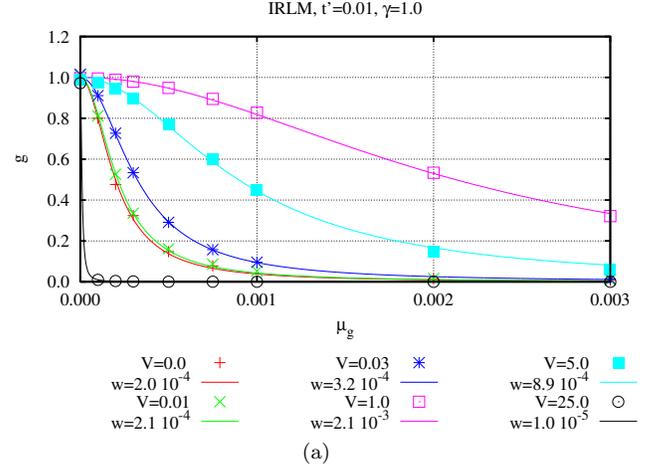}}
    \hspace{0.02\textwidth}
    \subfigure[]{\label{G_0.03}
    \includegraphics[width=\graphiclength]{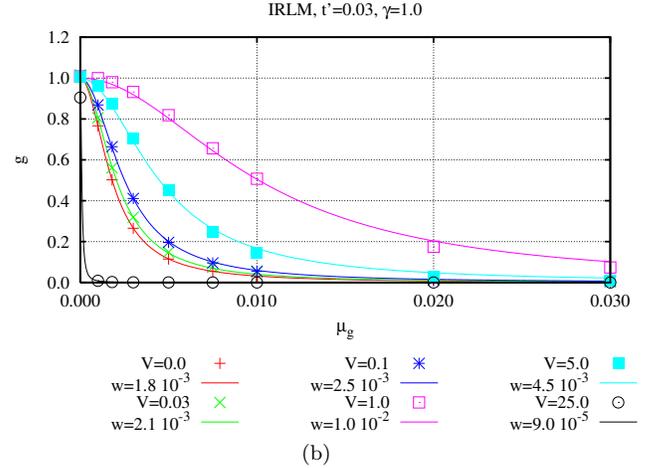}}
    \caption{\ColorOnline Conductance versus gate potential for the interacting resonant level
    model for a contact hopping of (a) $t'=0.01$ and (b) $t'=0.03$ and
    contact interaction ranging from zero to 25. To each set of DMRG data a Lorentzian of half width $2$w has been added as a guide to the eye.
    The leads are
    described with a cosine band between $\pm2$ such that the Fermi velocity is $v_F=2$.
    In contrast to intradot interaction the contact
    interaction enhances the conductance and shows a nonmonotonic behavior versus
    contact interaction.}\label{Fig: All G-1}
\end{center}
\end{figure}

\section{Results}
The aim of this work is to study the effect of contact interaction.
It is known from previous work
\cite{Bohr_Schmitteckert_Woelfle:2006} that strong repulsive interactions
within the nanostructure lead to suppression of the transport off resonance due to
the formation of a density-wave-like state on the dot.
\begin{figure}[tb]
\begin{center}
    \setlength{\graphiclength}{0.475\textwidth}
    \includegraphics[width=\graphiclength]{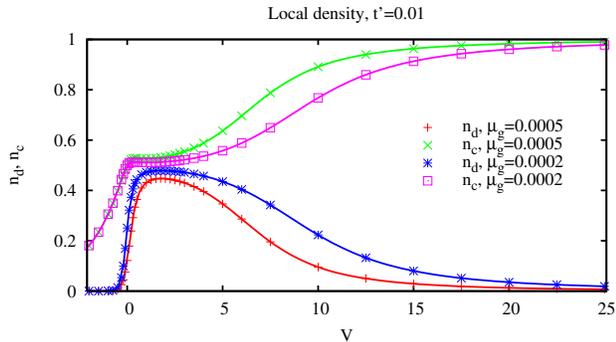}
    \caption{\ColorOnline Site occupation $n_d$ of the resonant level 
            and $n_c$ of the real-space sites attached to the level
            vs the link interaction in the IRLM for t'=0.01
            and two different gate voltages.}\label{Fig:SiteOccupation}
\end{center}
\end{figure}

In Fig.~\ref{Fig: All G-1} we show results for the conductance
versus gate potential for different couplings to the leads and
different contact interactions for the IRLM ($\gamma=1$). The
calculations have been performed with typically 130 sites in total,
$M_E=10$ real-space sites and 120 momentum-space sites. Due
to the symmetry of the band, we use a discretization that is
symmetric around $\epsilon_F=0$, and further use identical
discretization of the two leads. To represent the ``large'' energy
span in the band we use 20 logarithmically scaled sites, and
thereafter use 10 linearly spaced sites to represent the low-energy
scale correctly. In the DMRG calculations presented we used at least
1300 states per block and 10 finite lattice sweeps.
To each set of DMRG results in Fig.~\ref{Fig: All G-1} is added a Lorentzian of
half width $2$w as a guide to the eye. 

As the interaction is turned
up the width of the resonance is increased far beyond the noninteracting result, up to an order of magnitude larger; e.g., for
$t'=0.01$ and $V=1$ the resonance width is increased by a factor of
10. However, for a larger interaction $V>v_F=2$, transport is
suppressed, and for very large interactions the width even becomes
smaller than the noninteracting resonance. A similar nonmonotonic
behavior is observed by Borda
\emph{et~al.}\cite{Borda_Vladar_Zawadowski:2007}~using a
perturbative calculation and is opposite to the one originally reported by
Mehta and Andrei\cite{Mehta_Andrei:2006} which, however, has been corrected in
an erratum.\cite{Mehta_Chao_Andrei:2007}
Where preceding work%
\cite{Borda_Vladar_Zawadowski:2007,Mehta_Andrei:2006,Mehta_Andrei:2007,Mehta_Chao_Andrei:2007}
failed to reach the unitary limit, we demonstrate that
indeed the resonant value remains unitary.

\begin{figure}[t]
\begin{center}
   \setlength{\graphiclength}{0.475\textwidth}
    \includegraphics[width=\graphiclength]{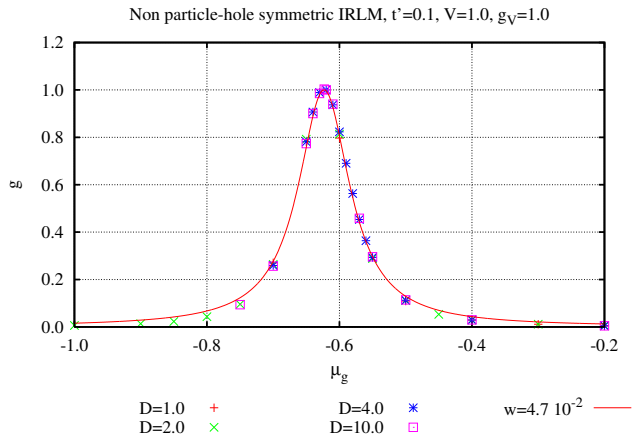}
    \caption{\ColorOnline Conductance versus gate potential for a single
    site nanostructure without particle-hole symmetry
    with a contact interaction of $V=1.0$ and a contact hopping of
    $t'=0.1$ for a linear band with cutoff parameter $D=1.0, 2.0, 4.0$, and $10.0$ and constant Fermi velocity, $v_F=2$. The
    conductance is independent of the cutoff. The solid line is a fit
    with a Lorentzian of half width w$=4.7~10^{-2}$.}\label{Fig: g-nph}
\end{center}
\end{figure}

Further by changing the bandwidth $D$ for linear bands we have
verified that the relevant energy scale is the \emph{Fermi velocity}
$v_{\text{F}}$ of the leads, while the bandwidth $D$ \emph{does not}
influence the conductance, as long as $D \gg V$; compare
Fig.~\ref{Fig: g-nph}.

Borda \emph{et~al.}\cite{Borda_Vladar_Zawadowski:2007} conclude in their work
that ``in the case of repulsive interaction the site next to the
occupied $d$ level is empty and thus that electron can easily
jump to the conduction band'', while for attractive interaction fermions accumulate
close to the impurity. From that reasoning we would expect an asymmetric 
conductance curve depending on whether the impurity is filled or depleted.
However, this would violate particle-hole symmetry of the model. 
In Fig.~\ref{Fig:SiteOccupation} we plot 
the site occupation $n_d$ of the resonant level and the averaged site occupation $n_c$
of the left and right real-space sites attached to the level. 
The occupations are plotted versus the contact link interaction
for the interacting resonant-level model, and for two different gate voltages.
The site occupation of the resonant level and the neighboring sites are
both enhanced by the repulsive interaction as long as interaction is in
the range that enhances the conductance. For stronger interaction the site occupancy 
of the resonant level is indeed reduced; however, this is the regime where the conductance is reduced.
We would like to remark that in the noninteracting case and for a weak contact, $t'\ll 1$,
the site occupations of the real-space sites in the leads changes only slightly with gate voltage
and are all very close to half filling. %
Thus it seems that the densities of the hybridizing lead levels are \emph{not} 
the determining quantity for the interaction-induced changes of transport properties.

\begin{figure}[t]
\begin{center}
    \setlength{\graphiclength}{0.475\textwidth}
    \includegraphics[width=\graphiclength]{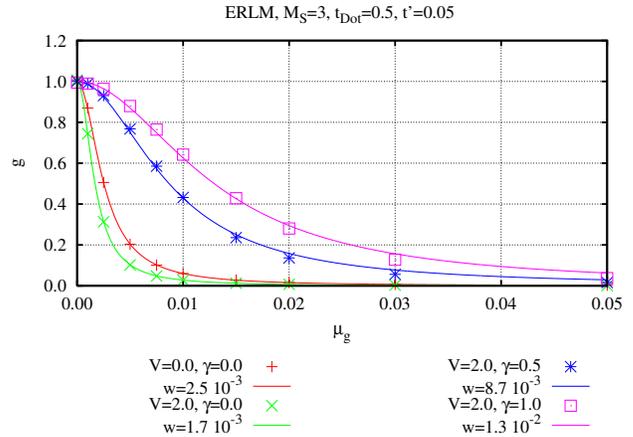}
    \caption{\ColorOnline Conductance versus gate potential for a resonant three
    site chain. To each set of DMRG results a Lorentzian of half width $2$w has been
    added as a guide to the eye.
    The leads are described by a cosine band between $\pm2$ such that $v_F=2$.
    The interdot interaction suppresses the transport
    while the contact interaction is seen to enhance the transport.}\label{Fig: g3}
\end{center}
\end{figure}

The strong renormalization of the resonance width and the non-monotonic behavior is, however, not specific to the IRLM. In Fig.~\ref{Fig: g3}
we show results for the center peak of a three-site
nanostructure. Without a contact interaction we find that the intradot
interaction $V=2.0 = 4 t_{\text{Dot}}$ leads to a
\emph{suppression} of the transport in agreement with previous
results.\cite{Bohr_Schmitteckert_Woelfle:2006} As in the single-level case already a small contact interaction increases again
the width of the resonance at zero gate potential. The enhancement
of the conductance by a contact interaction is stronger than the
corresponding suppression by the intradot interaction. Therefore we
conjecture that the enhancement of conductance due to the contact 
interaction is a universal feature, which should also be present in
other systems. These findings may also be relevant for disordered
structures, where repulsive interaction was found to enhance
transport in the case of strong
disorder.\cite{Molina_Schmitteckert_Weinmann_Jalabert_Ingold_Pichard:2004}

Finally we have considered a non-particle-hole-symmetric IRLM to
address the question of parameter renormalization versus bandwidth
cutoff. The non-particle-hole-symmetric model is defined by
replacing the $ (\hat{n}_j - \frac 12 )$ terms in $H_{RS}$ by $\hat{n}_j$.
The results are shown in Fig.~\ref{Fig: g-nph}. It is clearly seen
from the calculation that varying the cutoff over an order of
magnitude does not change the resonance, providing the interaction
is \emph{not} cut off by the band. Neither the position nor the
width of the resonance peak is influenced by the change of the
cutoff $D$, which is in contrast to the the renormalization group
flow that follows from the nonequilibrium Bethe
ansatz.\cite{Mehta_Andrei:2006} There, all transport quantities
depend on the cutoff $D$ and the conductance changes with the
cutoff. While it is often difficult to compare a field theoretical
model, like the IRLM of Mehta and Andrei, with a lattice
model, we can at least conclude that the RG flow found in their work
is absent in our model with regularized (tight binding) leads and that the relevant energy scale is the Fermi velocity.

\section{Summary}
A normal paradigm in transport calculations is to make a principal
division between transport region, the nanostructure or ``molecule'', and leads, where
all correlation effects are excluded from the leads.

In this work we have investigated the influence of an interaction on
the contact between a nanostructure and the leads in a simple
tight binding model. Using the nonperturbative DMRG method to evaluate the linear conductance we have
demonstrated that a contact interaction significantly influences
the transport properties. A repulsive interaction smaller or comparable to
the Fermi velocity in the leads  enhances the conductance, while
a large interaction leads to a suppression of the conductance. Our work
shows that even a slight spread of the interaction on the contacts
influences the transport strongly. This demonstrates that particular
care should be taken in treating the contacts correctly, especially
regarding the interaction.
\acknowledgments %
D.B.~acknowledges support from the HPC-EUROPA under Project No.~RII3-CT-2003-506079,
supported by the European Commission.
This work also profited from Project 710 of the Landesstiftung Baden-W\"{u}rttemberg and partial support through project B2.10 of the DFG Center for Functional Nanostructures.
Parts of the computations were performed on the XC1 and XC2 at the SSC Karlsruhe.
The discussion of the the site occupation is attributed to an anonymous referee.

\end{document}